%% file: main.tex
\documentclass[aps, prd, superscriptaddress, nofootinbib, preprintnumbers]{revtex4-2}
\usepackage[colorlinks=true]{hyperref}
\usepackage{xcolor}
\hypersetup{colorlinks=true, citecolor=blue, urlcolor=blue, linkcolor=blue}
\usepackage{amsmath,amssymb}
\usepackage[final]{graphicx}
\usepackage{subcaption}
\usepackage[belowskip=0.5pt,aboveskip=0.5pt]{caption}
\usepackage{float}
\usepackage{adjustbox}
\usepackage{cancel,soul,ulem}

\begin{document}

\title{Effects of Born-Infeld Electrodynamics on Chiral Symmetry Restoration and Meson Susceptibilities in Holographic QCD}

\author{Hiwa A. Ahmed}
\thanks{{\tt hiwa.ahmed@chu.edu.iq}}
\affiliation{Charmo Center for Research, Training, and Consultancy, Charmo University, 46023, Chamchamal, Sulaymaniyah, Iraq}

\author{Peshwaz A. Abdoul}
\thanks{{\tt peshwaz.abdoul@chu.edu.iq}}
\affiliation{Department of Physics, College of Science, Charmo University, 46023, Chamchamal, Sulaymaniyah, Iraq}

\date{\today}

\begin{abstract}
Within a holographic QCD framework, we numerically investigate chiral symmetry breaking and the associated phase transition at finite temperature and chemical potential. The model is constructed on a nonlinear charged Born-Infeld black hole background. The chiral condensate, extracted from the asymptotic behavior of the bulk scalar field, serves as the primary order parameter. At zero chemical potential, we find a chiral crossover transition for physical quark masses with a pseudocritical temperature of $T_{\rm pc}=0.1477$ GeV. In the chiral limit, the transition becomes first-order with a critical temperature of $T_{\rm c}=0.1337$ GeV. A critical strange quark mass of $m_s=37$ MeV, at zero light quark mass, separates first- and second-order transition regions. For finite chemical potential ($\mu$) and a physical strange mass ($m_s=95$ MeV) with massless light quarks, the transition remains second-order, with $T_c$ decreasing as $\mu$ increases. These results are further supported by the behavior of meson susceptibilities $(\chi_{\pi}-\chi_{\sigma})$, which exhibit a rapid thermal decay and convergence across the phase boundary. Introducing the Born-Infeld parameter $\beta$ shifts the second-order phase boundary to higher temperatures for smaller $\beta$—stabilizing the chirally broken phase—but does not alter the transition order or introduce a critical endpoint within the studied range. Our findings are consistent with previous soft-wall model studies and highlight the significant role of nonlinear bulk electrodynamics in modifying the chiral phase diagram.
\end{abstract}


\maketitle

\section{Introduction}

The exploration of the Quantum Chromodynamic (QCD) phase diagram remains a cornerstone of modern nuclear physics, as researchers strive to elucidate the behavior of quarks and gluons under extreme thermal and dense conditions. At low energy scales, QCD is characterized by color confinement \cite{Gross:2022hyw} and spontaneous chiral symmetry breaking \cite{Goldstone:1962es,Baker:1962zz}. However, as temperature ($T$) or chemical potential ($\mu$) increases QCD is expected to undergo a transition toward a deconfined, chirally symmetric state.

Understanding this transition remains a significant challenge. Lattice QCD (LQCD) provides accurate results for the chiral transition at zero chemical potential~\cite{Brown:1990ev,Aoki:2012yj,Ding:2015pmg,Bazavov:2017xul,HotQCD:2019xnw,Ding:2019fzc,Kuramashi:2020meg,Dini:2021hug}, but encounters the sign problem at finite $\mu$ and cannot reliably predict behavior at high densities \cite{Gattringer:2016kco}. This limitation necessitates the use of non-perturbative effective models. A prominent framework in this regard is the Anti-de Sitter/Conformal Field Theory (AdS/CFT) correspondence, or gauge/gravity duality \cite{Maldacena:1997re,Witten:1998qj}. Within holographic QCD, models are generally divided into top-down and bottom-up approaches. The former derives QCD-like properties from consistent string theory constructions \cite{Karch:2002sh,Sakai:2004cn}, while the latter constructs higher-dimensional gravitational duals designed to reproduce established QCD phenomenology \cite{Erlich:2005qh,DaRold:2005mxj,Karch:2006pv,Ghoroku:2005vt,Gubser:2008ny,Gubser:2008yx,Gursoy:2007cb,Gursoy:2007er}. A breakthrough in this field was the development of the soft-wall AdS/QCD model \cite{Karch:2006pv}. By introducing a specific dilaton profile, this model successfully captures the linear Regge trajectories of meson spectra and describes the spontaneous breaking of chiral symmetry. Consequently, the soft-wall framework has been extensively employed to map the phase structure across various flavor systems and thermodynamic regimes \cite{Colangelo:2011sr,Cui:2013zha,Bartz:2016ufc,Chelabi:2015cwn,Chelabi:2015gpc,Fang:2015ytf,Fang:2016nfj,Li:2016smq,Bartz:2017jku,Ahmed:2024rbj}.

Despite its successes, the standard soft-wall model typically relies on a linear Maxwell action for the bulk $U(1)$ gauge field to represent the chemical potential. While the Einstein-Maxwell framework provides a robust baseline for finite-density studies \cite{Cai:2012xh,Bohra:2020qom,Ballon-Bayona:2020xls}, the Maxwell approximation is fundamentally limited to low-energy regimes. To accurately model extreme physical conditions, nonlinear corrections to the electromagnetic sector become indispensable. Among these, the Born-Infeld (BI) theory stands out as a uniquely motivated extension. Originally proposed to regularize the divergent self-energy of point charges through a maximum field strength \cite{Born:1934gh}, BI electrodynamics emerges naturally as the low-energy effective theory of D-branes and open strings \cite{Fradkin:1985qd,Tseytlin:1986ti,Gibbons:2001gy}. Given that the gauge/gravity duality is rooted in string theory, adopting a BI background ensures a more consistent UV-complete description of bulk dynamics. In this work, we employ a BI black hole background, wherein the nonlinear parameter $\beta$ introduces a new physical scale that modifies gauge field behavior at short distances and high densities.

Furthermore, the coupling of BI electrodynamics to gravity introduces novel black hole solutions with thermodynamic properties and extremal structures that deviate significantly from the standard Reissner-Nordstrom (RN) case \cite{Dey:2004yt,Cai:2004eh}. In a holographic context, these modifications allow for a more subtle investigation of the chemical potential, as the BI action incorporates all-order higher-derivative corrections of the gauge field. This complexity is essential for capturing potential physical effects in the dual boundary theory that are otherwise invisible in linear Maxwell form descriptions. By utilizing the BI parameter $\beta$ as a tunable scale for nonlinearity, we can systematically probe the stability of the chirally broken phase and determine how short-distance modifications to the bulk electrodynamics influence the pseudocritical temperature and the order of the chiral transition. This approach not only evaluates the robustness of previous soft-wall models but also highlights the decisive role of nonlinear bulk dynamics in shaping the holographic QCD phase diagram.

In this paper, we investigate the impact of BI nonlinearity on the chiral phase boundary through two primary lenses: the chiral condensate, acting as the fundamental order parameter, and the scalar/pseudoscalar susceptibilities. The difference $(\chi_{\pi} - \chi_{\sigma})$ is utilized to pinpoint the temperature at which the chiral gap closes. Our numerical results demonstrate that nonlinear electrodynamics serves to stabilize the chirally broken phase; specifically, at high chemical potential, decreasing the BI parameter $\beta$ consistently elevates the critical temperature $T_c$ relative to the standard RN limit.

The paper is organized as follows. Firstly, we will provide a brief overview of the BI black hole in AdS in section \ref{sectionII}. Then, sections \ref{sectionIII} and \ref{sectionIV} detail the construction of the soft-wall model and the formulation of meson susceptibilities within this background. Section \ref{sectionV} presents the numerical computation of the vacuum expectation value. The core findings regarding the restoration of chiral symmetry—analyzed via the condensate and susceptibility differences at finite $T$ and $\mu$—are discussed in Sections \ref{sectionVI}, \ref{sectionVII}, and \ref{sectionVIII}. Finally, we present our conclusions in Section \ref{sectionIX}.

\section{Born-Infeld black holes in A\lowercase{d}S}
\label{sectionII}

The standard black holes follow Maxwell's electrodynamics, in which the electric field obeys the inverse square law $E \propto 1/r^2$ everywhere in space. At the origin, $r=0$, the field strength and self-energy of the charge become infinite. In contrast, the Born-Infeld black holes introduce an upper limit to the field strength through a nonlinearity parameter $\beta$. Far from the black hole, where the field is weak, the solutions reduce to the standard linear Maxwell's electrodynamics. The effect of the nonlinearity becomes important near the horizon and singularity, where the field is strong, as it smooths out the self-energy and ensures that the electric field remains finite at the origin. In this work, we employ the same action integral and conventions (with $16\pi G = 1$) as in \cite{Cai:2004eh}. We confine our analysis to a spacetime dimension of $n+1=5$ ($n=4$) and assume the planar case with zero curvature $k=0$. The action integral is expressed as follows:
\begin{equation}
	S=\int d^{5}x\,\sqrt{-g}\left(R-2\Lambda+L(F)\right), 
 \label{eq:action} 
\end{equation} 
where the Born-Infeld Lagrangian is given by 

\begin{equation}
L(F)=4\beta^{2}\left(1-\sqrt{1+\frac{F_{\mu\nu}F^{\mu\nu}}{2\beta^{2}}}\right), 
\label{eq:borninfeld}
\end{equation} 
with $\beta$ representing the Born-Infeld parameter. In the limit as $\beta \to \infty$, we retrieve the linear electrodynamic case, expressed as 

\begin{equation} 
	L(F) \approx - F_{\mu\nu}F^{\mu\nu} + ..... 
 \label{eq:borninfelde}
\end{equation}

Varying Eq. \eqref{eq:action} gives the equations of motion for the electromagnetic field and the Einstein field equations for the gravitational metric:

	\begin{align}
		\partial_{\mu}\!\left(\frac{\sqrt{-g}\,F^{\mu\nu}}{\sqrt{1+\dfrac{F^{2}}{2\beta^{2}}}}\right)&=0,
  \label{eq:EM}
	\end{align}
	and
	\begin{align}
		R_{\mu\nu}-\tfrac{1}{2}R g_{\mu\nu}+\Lambda g_{\mu\nu}&=\tfrac{1}{2}g_{\mu\nu}L(F)+\frac{2F_{\mu\alpha}F^{\alpha}{}_{\nu}}{\sqrt{1+\dfrac{F^{2}}{2\beta^{2}}}},
  \label{eq:Einstein}
	\end{align}
where $F^{2}=F_{\rho\sigma}F^{\rho\sigma}$.

\subsection{Electromagnetic field solution and Chemical Potential}

We consider a  restricted case of a fully electrostatic configuration, corresponding to $F_{tr}(r)$ being the only non-zero component of the field tensor. After this simplification, the field equation, Eq. \eqref{eq:EM}, is integrated to obtain

\begin{equation}
	F^{rt}(r)=\frac{\sqrt{6}\,\beta q}{\sqrt{2\beta^{2}r^{6}+6q^{2}}}.
 \label{eq:Frt_n4}
\end{equation}

Here, $q$ is the integration constant. The chemical potential is defined as the electrostatic potential difference between the boundary at $r=\infty$ and the horizon at $r=r_{h}$, i.e. $\mu = A_{t}(\infty) - A_{t}(r_h)$. The electrostatic potential $A_{t}(r)$ satisfies $F_{tr}=-\partial_{r}A_{t}$, resulting in the following:

\begin{equation}
	A_{t}(r)=\frac{\sqrt{3}q}{2r^{2}}\;{}_2F_{1}\Bigl(\tfrac{1}{3},\tfrac{1}{2};\tfrac{4}{3};-\frac{3q^{2}}{\beta^{2}r^{6}}\Bigr).
 \label{eq:At}
\end{equation}

By applying the boundary condition $A_{t}(r_h)=0$, the chemical potential is given by 

\begin{equation}
\mu = \frac{\sqrt{3}q}{2r_{h}^{2}}\;{}_2F_{1}\Bigl(\tfrac{1}{3},\tfrac{1}{2};\tfrac{4}{3};-\frac{3q^{2}}{\beta^{2}r_{h}^{6}}\Bigr).
\label{eq:mu_n4}
\end{equation}

Finally, we rewrite this in terms of holographic coordinate $z=1/r$ ($z_{h}=1/r_{h}$) and by employing the relation $Q = \frac{q}{r_{h}^{3}} = q z_{h}^{3}$, the chemical potential reads

	\begin{equation}
		\mu=\frac{\sqrt{3}}{2} \; \frac{Q}{z_{h}}
		\;{}_2F_{1}\Bigl(\tfrac{1}{3},\tfrac{1}{2};\tfrac{4}{3};-\frac{3Q^{2}}{\beta^{2}}\Bigr).
  \label{eq:mu_z}
	\end{equation}
 
By expanding this in the limit $\beta \to \infty$, retaining the first three terms, we obtain the following:

\begin{eqnarray}
	\mu \approx \frac{\sqrt{3} Q}{2 z_{h}} - \frac{3 \sqrt{3} Q^{3}}{16 z_{h} \beta^{2}} + \frac{27 \sqrt{3} Q^{5}}{112 z_{h} \beta^{4}} + \mathcal{O}\left(\frac{1}{\beta^6}\right).
\end{eqnarray}

\subsection{Einstein field equations, metric function and temperature}

By setting the spacetime dimension to $ n+1=5$ and assuming zero curvature ( $k=0$) with an AdS radius of $ \ell=1 $, the spacetime metric takes  the following form:
  
\begin{eqnarray} 
	ds^{2} = -f(r) dt^{2} + \frac{dr^{2}}{f(r)} + r^{2} (dx_{i}^2),
 \label{eq: metric}
\end{eqnarray}
where  $i = 1$, $2$, $3$.
With this arrangement, the Einstein field equations \ref{eq:Einstein} are subsequently solved to derive the metric function. In the context of the holographic coordinate $z=1/r$, this is expressed as

\begin{equation}
f(z)=-m z^{2} + 
	\left(\frac{\beta^{2}}{3}+1\right) \frac{1}{z^{2}}
	-\frac{\beta^2}{3z^{2}}\sqrt{1+ \frac{3q^{2} z^{6}}{\beta^{2}}}+
	\frac{3}{2}q^{2}z^{4}\,{}_2F_{1}\Bigl(\tfrac{1}{3},\tfrac{1}{2};\tfrac{4}{3};-\frac{3q^{2} z^{6}}{\beta^{2}}\Bigr).
 \label{eq:V_n4}
\end{equation}

The Hawking temperature is defined by $T=f'(z_{h})/(4\pi)$. After expressing the mass-related integration constant $m$ in terms of the horizon radius $z_{h}$ using the condition $f(z_{h})=0$, this reads:
	\begin{equation}
		T=\frac{1}{\pi z_{h}}
		\left[ 1 + \frac{\beta^{2}}{3} 
		\left(1- \sqrt{1 + \frac{3Q^2}{\beta^{2}}}\right) \right].
  \label{eq:T_z}
	\end{equation}
 
By applying the binomial expansion for the square root term, assuming $\beta >> 1$, we arrive at the following expression:

\begin{eqnarray}
T \approx \frac{1}{\pi z_{h}} - \frac{Q^2}{2\pi z_{h}} + \frac{3Q^4}{8\pi z_{h}\beta^2} + \mathcal{O}\left(\frac{1}{\beta^4}\right).
\end{eqnarray}

Furthermore, in the limit $\beta \to \infty$ corresponding to the linear Maxwell electrodynamics, the above equation reduces to the Hawking temperature for an RN black hole:
\begin{eqnarray}
	T = \frac{1}{\pi z_{h}} - \frac{Q^2}{2\pi z_{h}}.
\end{eqnarray}
Finally, for an uncharged black hole with $Q=0$, the Schwarzschild solution is obtained, which reads 
\begin{eqnarray}
	T = \frac{1}{\pi z_{h}}.
\end{eqnarray}

\section{Soft-Wall Model in a Born-Infeld Black Hole Background}
\label{sectionIII}

This section investigates the soft-wall holographic QCD model within the geometrical framework established previously. Adopting a bottom-up approach, we consider a five-dimensional gauge theory endowed with $U(N_f)_L\times U(N_f)_R$ chiral symmetry, where $N_f$ denotes the number of quark flavors \cite{Erlich:2005qh,DaRold:2005mxj,Karch:2006pv}. In this work, we study the behavior of the chiral phase transition at finite temperature and chemical potential, for which we only need to have the chiral symmetry breaking, which is implemented via a bulk scalar field $X$, transforming in the bifundamental representation of the chiral symmetry group. Conformal invariance is broken by a dilaton background field $\Phi(z)$, which serves as a soft infrared (IR) cutoff \cite{Karch:2006pv}.

The five-dimensional action revealing the chiral symmetry breaking in the soft-wall model is given by:
\begin{equation}
\begin{aligned}
S_{M} &= \int d^{5} x \sqrt{g} \, e^{-\Phi} \Bigg\{ \operatorname{Tr}\left[ |D^{M} X|^{2} - V(X) \right] - \gamma \, \mathrm{Re}(\det[X]) \Bigg\},
\end{aligned}
\label{action}
\end{equation}
where $g \equiv \det(g_{MN}) = e^{5A_s(z)} = z^{-5}$ (for which $A_{s}(z)= - \log(z/ \ell)$) is the determinant of the BI metric as defined in the previous section, and the covariant derivative is reduced to a normal derivative as $D^{M} X=\partial^{M} X$ because we do not consider the gauge field here to couple the scalar field. The parameter $\gamma$ governs the strength of the determinant term, $\mathrm{Re}(\det[X])$, which is anomalous under $U(1)_A$ but preserves $SU(N_f)_L\times SU(N_f)_R$ symmetry. The scalar potential $V(X)$ is given as follows

\begin{equation}
V(X)= m_{5}^{2} \operatorname{Tr}\left( X^{\dagger} X\right) + \lambda \operatorname{Tr}\left( |X|^{4}\right),
\label{potential}
\end{equation}
where the five-dimensional mass, from the AdS/CFT correspondence, is $m_{5}^{2} = (\Delta - p)(\Delta + p - 4) = -3$ for an operator of scaling dimension $\Delta=3$ and $p=0$. The positive dimensionless coupling $\lambda$ ensures a non-vanishing quark condensate in the chiral limit ($m_q \to 0$); without it, the condensate would vanish accidentally \cite{Gherghetta:2009ac}.

To establish a model that can obtain both a consistent meson spectrum and a satisfactory description of the chiral phase transition within the soft-wall model framework, a specific dilaton profile that respects both IR and UV behavior is necessary. Such a dilaton profile is given in Refs. \cite{Chelabi:2015cwn,Chelabi:2015gpc} as

    \begin{equation}
        \Phi(z) = - \mu_{1}^{2}z^2  + (\mu_{1}^{2} + \mu_{g}^{2}) z^{2} \tanh(\mu_{2}^{2} z^{2}).
    \end{equation}

To analyze chiral symmetry breaking, we consider the vacuum expectation value of the scalar field $X$, which assumes the form:
\begin{equation}
X(z) \equiv X_0(z) = \mathrm{diag} \left( \frac{\chi_l(z)}{2}, \frac{\chi_l(z)}{2}, \frac{\chi_s(z)}{2}\right),
\label{chivev}
\end{equation}
where isospin symmetry ($m_u = m_d \equiv m_l$) is imposed, leading to degenerate light quark components $\chi_u(z) = \chi_d(z) \equiv \chi_l(z)$, while $\chi_s(z)$ corresponds to the strange quark.

Substituting the vacuum expectation values, $\chi_l$ and $\chi_s$, in the action \eqref{action}, it reduces to

\begin{equation}
\begin{aligned}
S\left[\chi_l,\chi_s\right] = \int d^5 x \sqrt{g} \, e^{-\Phi} \Bigg\{ g^{zz} \left( \frac{1}{2} \chi_l^{\prime 2} + \frac{1}{4} \chi_s^{\prime 2} \right) - \left( \frac{1}{2} m_{5}^{2} \chi_l^2 + \frac{1}{4} m_{5}^{2} \chi_s^2 + \frac{\lambda}{8} \chi_l^4 + \frac{\lambda}{16} \chi_s^4 + \frac{\gamma}{8} \chi_l^2 \chi_s \right) \Bigg\},
\end{aligned}
\label{actionchi}
\end{equation}
where $g^{zz} = -e^{-2A_s(z)} f(z)$ is the relevant component of the inverse metric.

\subsection{Equations of Motion}

Varying the effective action \eqref{actionchi} yields the equations of motion (EOMs) for the fields $\chi_{f}(z)$ ($f = l, s$):
\begin{equation}
\begin{gathered}
    \chi_l^{\prime \prime} + \left( 3 A_s^{\prime} - \Phi^{\prime} + \frac{f^{\prime}}{f} \right) \chi_l^{\prime} - \frac{e^{2 A_s}}{f} \left( m_{5}^{2} + \frac{\gamma}{4} \chi_s + \frac{\lambda}{2} \chi_l^2 \right) \chi_l = 0, \\
    \chi_s^{\prime \prime} + \left( 3 A_s^{\prime} - \Phi^{\prime} + \frac{f^{\prime}}{f} \right) \chi_s^{\prime} - \frac{e^{2 A_s}}{f} \left( m_{5}^{2} \chi_s + \frac{\gamma}{4} \chi_l^2 + \frac{\lambda}{2} \chi_s^3 \right) = 0,
\end{gathered}
\label{chieom3}
\end{equation}
where the prime ($^\prime$) denotes a derivative with respect to the holographic coordinate $z$.

The asymptotic behavior of $\chi_l(z)$ and $\chi_s(z)$ in the ultraviolet (UV, $z \to 0$) and infrared (IR, $z \to z_h$) regions are derived as follows:

\begin{equation}
\begin{gathered}
    \chi_{l}(z\to 0) = m_{l} \zeta z - \frac{\gamma}{4} m_{l} m_{s} \zeta^{2} z^2 + \frac{m_{l} \zeta}{4} \left( -4 \mu_{1}^{2} + m_{l}^{2} \zeta^{2} \lambda - \frac{\gamma^{2}}{8} (m_{l}^{2} \zeta^2 + m_{s}^{2} \zeta^2) \right) z^{3} \log z + \frac{\sigma_l}{\zeta} z^{3} + \dots, \\
    \chi_{l}(z\to z_h) = a_{l,0} + \frac{a_{l,0} \beta^2 ( -2 a_{l,0}^{2} \lambda + 12 - \gamma a_{s,0})}{2 z_h (-8 \beta^2 +4 \beta^2 Q^2 - 3 Q^4)} (z_h - z) + \mathcal{O}[(z_h - z)^{2}], \\
    \chi_{s}(z\to 0) = m_{s} \zeta z - \frac{\gamma}{4} m_{l}^{2} \zeta^{2} z^2 + \frac{1}{4} \left( -4 \mu_{1}^{2} m_{s} \zeta + m_{s}^{3} \zeta^{3} \lambda - \frac{\gamma^{2}}{4} m_{l}^{2} m_{s} \zeta^3 \right) z^{3} \log z + \frac{\sigma_s}{\zeta} z^{3} + \dots, \\
    \chi_{s}(z\to z_h) = a_{s,0} + \frac{\beta^2 ( -2 a_{s,0}^{3} \lambda + 12 a_{s,0} - \gamma a_{l,0}^{2} )}{2 z_h (-8 \beta^2 +4 \beta^2 Q^2 - 3 Q^4)} (z_h - z) + \mathcal{O}[(z_h - z)^{2}],
\end{gathered}
\label{UV3l1}
\end{equation}

\noindent here, $m_l$ and $m_s$ are the light and strange quark masses, $\sigma_l$ and $\sigma_s$ are the corresponding chiral condensates, and $\zeta = \sqrt{N_c}/(2\pi)$ is a normalization constant ensuring consistent large-$N_c$ scaling for the $\chi$ field. The constants $a_{l,0}$ and $a_{s,0}$ are IR integration constants.

%%%%%%%%%%%%%%%%%%%%%%%%%%%%%%%%%%%%%%%%%%%%%%%%%%%%%%%%%%%%%%%%%%%%%%%%%%%

\section{meson susceptibility}
\label{sectionIV}

An alternative technique to study the chiral symmetry breaking and/or restoration is through the meson susceptibilities, where for the chiral partners, the susceptibilities must be degenerate at the chiral restoration region \cite{Ahmed:2026qzw}. The meson susceptibility is defined by the two-point correlation function at zero momentum. To study the susceptibility in holographic QCD, we need to consider a five-dimensional gauge theory with $U(N_f)_L\times U(N_f)_R$ symmetry, where $N_f$ is the number of quark flavors \cite{Erlich:2005qh,DaRold:2005mxj,Karch:2006pv} (here the number of flavors is chosen to be three, $N_f=3$). The model includes bulk left- ($L_M$) and right- ($R_M$) handed gauge fields and a bulk scalar field $X$. The five-dimensional action is given by
\begin{equation}
\begin{aligned}
S_{M} &=\int d^{5} x \sqrt{g} e^{-\Phi}\left\{ \operatorname{Tr}\left[\left(D^{M} X\right)^{\dagger}\left(D_{M} X\right)-V(X) -\frac{1}{4 g_{5}^{2}}\left(L^{M N} L_{M N}+R^{M N} R_{M N}\right)\right]- \gamma Re(det[X])
\right\},
\end{aligned}
\label{actions}
\end{equation}
where $D^{M}$ is the covariant derivative which is $D^{M} X=\partial^{M} X - i L^{M} X + i X R^{M}$, and  $L_{MN}=\partial_M L_N -\partial_N L_M -i[L_M,L_N]$ and $R_{MN}=\partial_M R_N -\partial_N R_M -i[R_M,R_N]$ are the left and right-handed field strength tensors, respectively. The gauge coupling $g_{5}$ is fixed by matching the vector current correlator to its QCD asymptotic form: $g_{5}^{2} = 12 \pi^{2} /N_{c}$.

For the scalar and pseudo-scalar mesons, the perturbations on the background have the following form,

\begin{equation}
X= (X_0 +S)  e^{2 i \pi},
\label{Xfield}
\end{equation}
with $S$ and $\pi$ are the scalar and pseudoscalar field fluctuations, respectively. 

\subsection{Scalar sector}

The scalar field is defined by a $3 \times 3$ matrix,

\begin{equation}
S=S^a t^a=\frac{1}{\sqrt{2}}\left(\begin{array}{ccc}
\frac{a_0}{\sqrt{2}}+\frac{\sigma_8}{\sqrt{6}}+\frac{\sigma_0}{\sqrt{3}} & a^{+} & \kappa^{+} \\
a^{-} & -\frac{a_0}{\sqrt{2}}+\frac{\sigma_8}{\sqrt{6}}+\frac{\sigma_0}{\sqrt{3}} & \kappa^0 \\
\kappa^{-} & \bar{\kappa}^0 & -\frac{2 \sigma_8}{\sqrt{6}}+\frac{\sigma_0}{\sqrt{3}}
\end{array}\right),
\end{equation}
where $t^{a}=\lambda_{a}/2$ $(a=0,1,...,8)$ are the generators of $U(3)$, where $\lambda_{a=1,...,8}/2$ are the Gell-Mann matrices with $\lambda_{0}=\sqrt{2/3} \mathbf{ 1}_{3 \times 3} $. The $\sigma_0$ and $\sigma_8$ are referred to as the admixtures of the $\sigma$ and $f_{0}(980)$.

Keeping only the scalar fluctuation in the action of Eq. \eqref{actions}, and expanding up to the second order, the action reduces to

\begin{equation}
S_{\mathrm{S}}=\frac{1}{2} \int d x^5 \sqrt{g} e^{-\Phi}\left[g^{\mu \nu} \partial_\mu S^{a} \partial_\nu S^{b} +g^{z z}\left(\partial_z S^{a}\right)\left(\partial_z S^{b}\right)-m_5^2 S^{a} S^{b}-M^{a,b}(z) S^{a} S^{b}\right] ,
\end{equation}
with the mass-like term is given by

\begin{equation}
\begin{gathered}
M^{a,b}(z)= \lambda M^{a,b}{s}(z) +2 \gamma M^{a,b}{det} (z),\\
M^{a,b}_{s}(z)=2 Tr \left( {t^{a},X_0}{t^{b},X_0} +2 t^{a} t^{b} X_0 X_0  \right),
\end{gathered}
\label{massscalar}
\end{equation}

For convenience, the system can be transformed from coordinate space (x) to momentum space (p) using the Fourier transformation $S^{a}(\mathrm{x}, z)=\frac{1}{(2 \pi)^4} \int d^4 \mathrm{p} e^{i \mathrm{p.x}} S^{a}(\mathrm{p}, z)$. In this section, we examine the two-point correlation function and susceptibility of the $\sigma$ meson in the scalar sector.

The mass-like parameter $M^{a,b}(z)$ in Eq. \eqref{massscalar} is non-diagonal, resulting in mixing between the singlet and octet states. To obtain the equations of motion for the physical states $\sigma$ and $f_0(980)$, we diagonalize the mass term $M^{a,b}(z)$ using an orthogonal transformation as follows \cite{Kawaguchi:2020qvg}:

\begin{equation}
\begin{aligned}
\tilde{S}^{i} & =O^{i a} S^{a}, \\
\tilde{M}^{i,j} & =O^{i a}\left(M^{a, b}\right) O^{b i} .
\end{aligned}
\end{equation}

The values of $M^{a,b}(z)$ for the singlet and octet states, including their mixing, are given by

\begin{equation}
\begin{aligned}
&   M^{0,0}= \lambda M^{0,0}{s} + \gamma M^{0,0}{det}=\frac{\lambda}{2}(2 \chi_{l}^{2} + \chi_{s}^{2}) + \frac{\gamma}{6} (2 \chi_{l} + \chi_{s}),\\
&   M^{8,8}= \lambda M^{8,8}{s} + \gamma M^{8,8}{det}=\frac{\lambda}{2}( \chi_{l}^{2} +2 \chi_{s}^{2}) - \frac{\gamma}{12} (4 \chi_{l} - \chi_{s}),\\
&   M^{0,8}= \lambda M^{0,8}{s} + \gamma M^{0,8}{det}=\frac{\lambda}{2}( \chi_{l}^{2} - \chi_{s}^{2}) - \frac{\gamma}{3\sqrt{2}} (4 \chi_{l} - \chi_{s}).\\
&
\end{aligned}
\end{equation}

The corresponding mass-like parameters for the physical states $\sigma$ and $f_{0}(980)$ are

\begin{equation}
\begin{aligned}
& M_{\sigma}(z)=M^{0,0}(z) \cos ^2 \theta_S+M^{8,8}(z) \sin ^2 \theta_S + 2 M^{0,8}(z) \cos \theta_S \sin \theta_S, \\
& M_{f_0(980)}(z)=M^{0,0}(z) \sin ^2 \theta_S+M^{8,8}(z) \cos ^2 \theta_S - 2 M^{0,8}(z) \cos \theta_S \sin \theta_S, \\
&
\end{aligned}
\end{equation}
where $\theta_S$ represents the scalar mixing angle,
\begin{equation}
\tan 2 \theta_S(z)=\frac{2 M^{0,8}(z)}{M^{0,0}(z)-M^{8,8}(z)} .
\end{equation}
Unlike in the conventional linear sigma model, the mixing angle $\theta_S$ here depends on the holographic coordinate $z$. The equation of motion for $\sigma$ can be derived from the action as follows:

\begin{equation}
S_{\mathrm{\sigma}}=\frac{1}{2} \int d x^5 \sqrt{g} e^{-\Phi}\left[g^{\mu \nu} \partial_\mu S \partial_\nu S+g^{z z}\left(\partial_z S\right)^2-m_5^2 S^2-M_{\sigma}(z) S^2\right] .
\end{equation}

\begin{equation}\label{pscalareq}
S^{\prime \prime}+\left(3 A^{\prime}+\frac{f^{\prime}}{f}-\Phi^{\prime}\right) S^{\prime}-\left(\frac{p^2}{f}+\frac{ m_5^2+ M_{\sigma}(z)}{ f} A^{\prime 2}\right) S=0
\end{equation}

The asymptotic solutions for $\sigma$ near the boundary and the horizon are as follows. Near the boundary, the solution is

\begin{equation}
S(z \rightarrow 0)=s_1 z + s_1 s_{\gamma} z^2     - \frac{1}{2}s_1 \left(2 \mu_1^2 -p^2 + s_{\gamma}^2 + s_{\lambda}\right) z^3 \log (z)+s_3 z^3+\mathcal{O}\left(z^4\right),
\end{equation}
where

\begin{equation}
\begin{aligned}
& s_{\gamma}= \gamma \zeta  \left( \frac{1}{12} (4 m_l-m_s) (\sin \theta_S(\epsilon))^2 -\frac{1}{6} (2 m_l+m_s) (\cos \theta_S(\epsilon))^2  + \frac{1}{3\sqrt{2}} (m_l-m_s) \sin \theta_S(\epsilon)  \cos \theta_S(\epsilon)      \right), \\
& s_{\lambda}= \lambda \zeta^2 \left(  \frac{1}{2} ( m_l^2 +2 m_s^2) (\sin \theta_S(\epsilon))^2 + \frac{1}{2} (2 m_l^2 + m_s^2) (\cos \theta_S(\epsilon))^2 + \sqrt{2} (m_l^2 - m_s^2) \sin \theta_S(\epsilon)  \cos \theta_S(\epsilon)     \right).\
&
\end{aligned}
\end{equation}

Near the horizon, the solution is

\begin{equation}
S\left(z \rightarrow z_h\right)=s_{h 0} - \frac{2 \beta^2 (-3 + M_{\sigma}(zh) + p^2 zh^2)}{(- \beta^2 + 4 \beta^2 Q^2 - 3 Q^4) z_h} s_{h 0}\left(z_h-z\right)+\mathcal{O}\left[\left(z_h-z\right)^2\right].
\end{equation}

By substituting the equation of motion into the action, the corresponding on-shell action is

\begin{equation}
S_{\mathrm{\sigma}}^{\text {on }}=-\left.\frac{1}{2} \int d^4 p f(z) S(-p, z) e^{3 A(z)-\Phi(z)} S^{\prime}(p, z)\right|_{z=\epsilon}^{z=z_h},
\end{equation}
In this context, $\epsilon$ serves as a UV cutoff to regularize the on-shell action. The two-point Green’s function for the scalar meson is determined by taking the second derivative of the on-shell action $S_{\mathrm{S}}^{\text {on }}$ with respect to the external source $J_S$. The two-point correlation function and the $\sigma$ susceptibility are expressed as follows:

\begin{equation}
G_{\mathrm{\sigma}}(p)=\left.\frac{\delta^2 S_{\mathrm{\sigma}}^{\mathrm{on}}}{\delta J_{\sigma}^* \delta J_{\sigma}}\right|_{z=\epsilon}=-\frac{4 s_3}{s_1} - 2 s{\gamma}^2  + \frac{1}{2} \left(2 \mu_1^2 -p^2 + s_{\gamma}^2 + s_{\lambda}\right).
\end{equation}

\begin{equation}
\chi_{\sigma} = -  \lim_{p^{2}\to 0} G_{\mathrm{\sigma}}(p) =\frac{4 s_3}{s_1} + 2 s_{\gamma}^2  - \frac{1}{2} \left(2 \mu_1^2+ s_{\gamma}^2 + s_{\lambda}\right).
\label{chisig}
\end{equation}

A comparison of Eq. \eqref{chisig} with the direct method for determining $\chi_{\sigma}$, which involves differentiating the chiral condensate with respect to the light quark mass, indicates that a normalization factor ($\zeta^2/4$) must be included in Eq. \eqref{chisig} (see the proof in Ref. \cite{Ahmed:2026qzw}).

\subsection{Pseudoscalar sector}

In this section, we examine the two-point correlation function and the susceptibilities of the pion meson. We introduce the pseudoscalar field as a small disturbance to the background, as shown in Eq. \eqref{Xfield}. Here, $\pi$ stands for the pseudoscalar field and is defined below:

\begin{equation}
\pi= \pi^a t^a=\frac{1}{\sqrt{2}}\left(\begin{array}{ccc}
\frac{\pi^0}{\sqrt{2}}+\frac{\eta_8}{\sqrt{6}}+\frac{\eta_0}{\sqrt{3}} & \pi^{+} & K^{+} \\
\pi^{-} & -\frac{\pi^0}{\sqrt{2}}+\frac{\eta_8}{\sqrt{6}}+\frac{\eta_0}{\sqrt{3}} & K^0 \\
K^{-} & \bar{K}^0 & -\frac{2 \eta_8}{\sqrt{6}}+\frac{\eta_0}{\sqrt{3}}
\end{array}\right).
\end{equation}

The axial-vector field also affects the pseudoscalar sector by interacting with the pion field. In the pseudoscalar channel, the pion field and the longitudinal part ($\varphi$) of the axial-vector field are linked. The fluctuation of the pion field is then given by

\begin{equation}
\begin{aligned}
S_\pi= & -\frac{1}{2 g_5{ }^2} \int d^5 x \sqrt{g} e^{-\Phi} \sum_{i=1}^3\left\{g^{\mu \nu} g^{z z} \partial_z \partial_\mu \varphi^i \partial_z \partial_\nu \varphi^i-g_5{ }^2 \chi_l^2\left(g^{\mu \nu} \partial_\mu \varphi^i \partial_\nu \varphi^i\right.\right. \\
& \left.\left.+g^{\mu \nu} \partial_\mu \pi^i \partial_\nu \pi^i+g^{z z}\left(\partial_z \pi^i\right)^2-2 g^{\mu \nu} \partial_\mu \varphi^i \partial_\nu \pi^i\right)\right\} .
\end{aligned}
\end{equation}

The EOMs for the pion field and the axial-vector field can be derived from the action as follows:
\begin{equation}
\begin{aligned}
\varphi^{\prime \prime}+\left(A^{\prime}+\frac{f^{\prime}}{f}-\Phi^{\prime}\right) \varphi^{\prime}-\frac{e^{2 A} g_5^2 \chi_l^2}{f}(\varphi-\pi) & =0, \\
\pi^{\prime \prime}+\left(3 A^{\prime}+\frac{f^{\prime}}{f}-\Phi^{\prime}+\frac{2 \chi_l^{\prime}}{\chi_l}\right) \pi^{\prime}+\frac{p^2}{f}(\varphi-\pi) & =0 .
\end{aligned}
\label{eom}
\end{equation}

The asymptotic solutions of the EOMs for the pion field at the boundary are given by \cite{Cao:2021tcr}:

\begin{equation}
\begin{aligned}
& \varphi(z \rightarrow 0)=c_f-\frac{1}{2} \zeta^2 g_5^2 m_l^2 \pi_0 z^2 \log (z)+\varphi_2 z^2+\mathcal{O}\left(z^3\right), \\
& \pi(z \rightarrow 0)=\pi_0+c_f + \frac{1}{2} \pi_0 p^2 z^2 \log (z)+\pi_2 z^2+\mathcal{O}\left(z^3\right),
\end{aligned}
\end{equation}
In this context, $c_f, \varphi_2, \pi_0$, and $\pi_2$ are integration constants. The external source $J_\pi$ of the pseudoscalar field corresponds to $\pi_0 m_q$ \cite{Ahmed:2026qzw}. As shown in \cite{Cao:2020ryx}, $c_f$ is a redundant parameter and can be set to zero for simplicity. The boundary conditions at the horizon are:

\begin{equation}
\begin{aligned}
& \varphi\left(z \rightarrow z_h\right)=-\frac{2 a_{l,0}^2 \beta^2 g_5^2 }{z_h (8 \beta^2 - 4 \beta^2 Q^2 + 3 Q^4)} \pi_{h 0}\left(z_h-z\right)+\mathcal{O}\left[\left(z-z_h\right)^2\right], \\
& \pi\left(z \rightarrow z_h\right)=\pi_{h 0}+\frac{2 \beta^2 p^2 z_h}{(8 \beta^2 - 4 \beta^2 Q^2 + 3 Q^4)} \pi_{h 0}\left(z_h-z\right)-\mathcal{O}\left[\left(z-z_h\right)^2\right] .
\end{aligned}
\end{equation}

$\pi_{h 0}$ is an additional integration constant. The on-shell action for the pion sector is given by

\begin{equation}
S_\pi^{\mathrm{on}}=-\left.\frac{1}{2 g_5^2} \int d^4 p e^{A-\Phi}\left[e^{2 A} g_5^2 f \chi^2 \pi(-p, z) \pi^{\prime}(p, z)+p^2 f \varphi(-p, z) \varphi^{\prime}(p, z)\right]\right|_{z=\epsilon} ^{z=z_h}.
\end{equation}

Equation \eqref{eom} leads to a first-order differential equation for the $\pi$ and $\varphi$ fields as

\begin{equation}
       p^{2} \partial_{z} \varphi + e^{2 A(z)} g_{5}^{2} \chi^{2} \partial_{z} \pi=0
\end{equation}

In holographic QCD, the fields are written in terms of the source and the bulk-to-boundary propagator. Near the boundary, $\pi$ and $\varphi$ are

\begin{equation}
\begin{aligned}
& \varphi(z \rightarrow 0)=\pi_0 \varphi_b=\pi_0 \left(-\frac{1}{2} \zeta^2 g_5^2 m_l^2 z^2 \log (z)+\frac{\varphi_2}{\pi_0} z^2+\mathcal{O}\left(z^3\right) \right), \\
& \pi(z \rightarrow 0)=\pi_0 \pi_b =\pi_0 \left(1+ \frac{1}{2} p^2 z^2 \log (z)+ \frac{\pi_2}{\pi_0} z^2+\mathcal{O}\left(z^3\right) \right).
\end{aligned}
\end{equation}

Here, $b$ stands for the bulk-to-boundary propagator. To find a solution for the pion field, we can use a perturbative approach in $m_{\pi}$ by setting $\varphi_b(z)=a_1(0,z)-1$ \cite{Erlich:2005qh}, where $a_1$ is the bulk-to-boundary propagator for the axial-vector field of the $a_1$ meson with $p^{2}=0$. The on-shell action for the pion is

\begin{equation}
S_\pi^{\mathrm{on}}=-\left.\frac{1}{2 g_5^2} \int d^4 p e^{A-\Phi} \pi_0 \pi_0 \left[- f m_{\pi}^{2} \pi_b(-p, z)  a_1^{\prime}(0, z) +p^2 f \varphi_b(-p, z) \varphi_b^{\prime}(p, z)\right]\right|_{z=\epsilon} ^{z=z_h},
\label{actionpi}
\end{equation}
where $m_{\pi}^2=-\bold{p}^2$ is the screening mass of the pion. Using the solution for the $a_1$ field and the ultraviolet boundary conditions for the bulk-to-boundary propagator of the axial-vector field, we find the two-point correlation function of the pion,

\begin{equation}
G_{\mathrm{\pi}}(p) =\left.\frac{\delta^2 S_{\mathrm{\pi}}^{\mathrm{on}}}{\delta J_{\pi}^* \delta J_{\pi}}\right|_{z=\epsilon} =\frac{m_{\pi}^{2} }{2 m_l^2 g_{5}^{2}} \left[\frac{2  a_{1,2}}{a_{1,0}} + \frac{1}{2} (\zeta^2 g_5^2 m_l^2 )   \right],
\end{equation}
with $a_{1,0}$, which corresponds to the source of the axial-vector field, and $a_{1,2}$ being the integration constants \cite{Ahmed:2026qzw}. Using the standard definition of susceptibility, the pion susceptibility is given by

\begin{equation}
\chi_{\pi} = - \lim_{p^{2}\to 0} G_{\mathrm{\pi}}(p) =- \frac{m_{\pi}^{2} }{2 m_l^2 g_{5}^{2}} \left[\frac{2  a_{1,2}}{a_{1,0}} + \frac{1}{2} \zeta^2 g_5^2 m_l^2   \right].
\label{susc}
\end{equation}

%%%%%%%%%%%%%%%%%%%%%%%%%%%%%%%%%%%%%%%%%%%%%%%%%%%%%%%%%%%%%%%%%%%%%%

\section{Numerical solutions of the vacuum expectation value}
\label{sectionV}

Within the holographic framework, the quark condensate in QCD is identified with the subleading term in the ultraviolet (UV) asymptotic expansion of the bulk scalar fields $\chi_f$ ($f=l,s$), which act as the order parameters for spontaneous chiral symmetry breaking. According to the AdS/CFT dictionary, in the near-boundary behavior of the field $\chi$ as given by Eq. \eqref{UV3l1}, the coefficient of the leading term, $m_f$, corresponds to the current quark mass (the source). The coefficient of the subleading term, $\sigma_f$, is identified with the chiral condensate $\langle \bar{q}q \rangle_f$ (the vacuum expectation value)~\cite{Erlich:2005qh,Karch:2006pv}.

Before moving to solve the equations of motion for the scalar field, we need to fix the parameters of the model. The model parameters are listed in Table \ref{tab:parameter}. These parameters are chosen to achieve a physical mass of the rho meson and a pseudocritical temperature of approximately $145$ MeV at zero chemical potential.

The coupled equations of motion in Eq.~\eqref{chieom3} are solved numerically using the shooting method, with boundary conditions given by Eqs.~\eqref{UV3l1}. The implementation of this method follows the procedure outlined in Refs.~\cite{Chelabi:2015cwn,Ahmed:2024rbj}. Numerical integration is performed from the near-horizon region at $z = z_h - \epsilon$ to the near-UV boundary at $z = \epsilon$.

\begin{figure}
\begin{subfigure}{0.49\textwidth}
  \centering
  \includegraphics[width=1\linewidth]{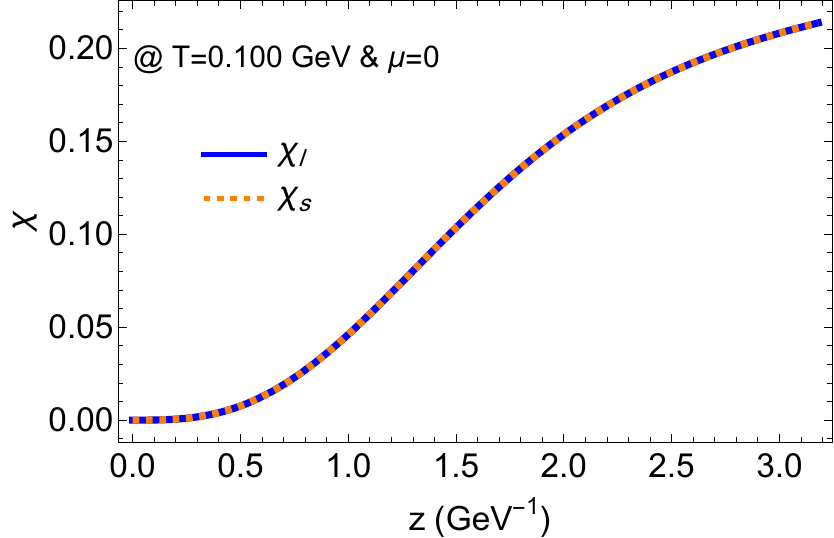} 
   \caption{ }
\label{chisolutionza}
\end{subfigure} 
\begin{subfigure}{0.49\textwidth}
  \centering
  \includegraphics[width=1\linewidth]{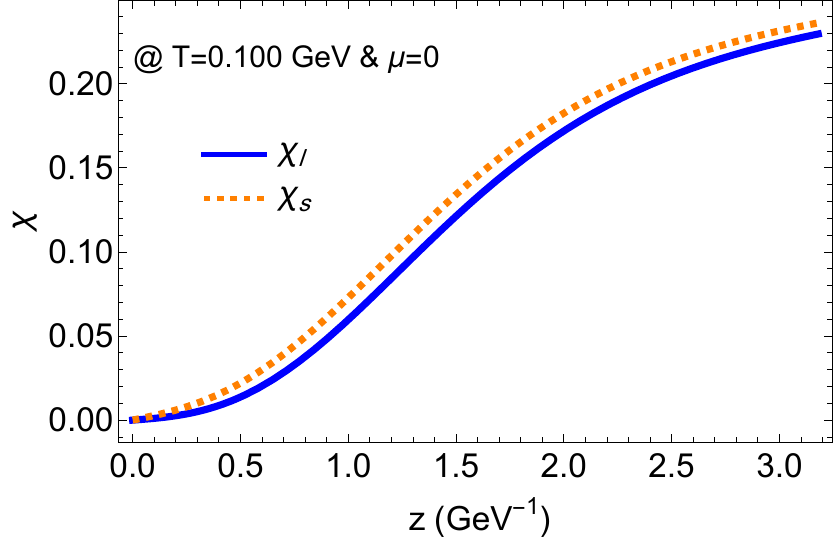}
 \caption{ }
\label{chisolutionzb}
\end{subfigure}%
\caption{Solutions of $\chi_l$, and $\chi_s$ as a function of the holographic coordinate $z$ at massless limit $m_{l}=m_{s}=0$ (a) and physical masses $m_l=3.5 $ MeV and $m_s=95$ MeV (b) for $T=0.100$ GeV and $\mu=0$. }
\label{chisolutionz}
\end{figure}

The resulting solutions for the scalar field $\chi_f$ are presented in Fig.~\ref{chisolutionz} for the quark masses at the chiral limit and physical masses. The calculations are carried out for the temperature, $T = 0.100~\mathrm{GeV}$ and chemical potential, $\mu=0$. As shown in Fig.~\ref{chisolutionz}, the numerical solutions satisfy the imposed boundary conditions and exhibit no singular behavior.

\begin{table} 
\center
\begin{tabular}{c c  c cc c c}
\hline
\hline
$\lambda = 80$  &      &   $\gamma = -25$ &  & $\mu_{g} = 0.43$ (GeV)    \\ 
$\mu_{1} =0.72$ (GeV)      &      & $\mu_{2} =0.176$ (GeV)      &      &  \\
     \hline
     \hline
\end{tabular}
\caption{The values of the parameters in the model.}
\label{tab:parameter}
\end{table}

The quark condensate $\sigma_f$ is extracted by matching the ultraviolet (UV) asymptotic behavior of the scalar field $\chi_f$ to the expression given in Eq.~\eqref{UV3l1}. In the massless quark limit, the flavor symmetry ensures identical solutions for all $\sigma_f$, yielding equal condensates $\sigma_l = \sigma_s$. At temperatures $T = 0.100~\mathrm{GeV}$, the corresponding condensate values are $\sigma_l = \sigma_s =0.264 ~(\mathrm{GeV})^3$. Considering the finite quark masses, one can see that the separation of $\chi_l$ and $\chi_s$, which results in the flavor symmetry breaking, as the solutions differ significantly across flavors. This is directly reflected in the computed condensates: $\sigma_l =0.272 ~(\mathrm{GeV})^3$, $\sigma_s =0.263 ~(\mathrm{GeV})^3$.

%%%%%%%%%%%%%%%%%%%%%%%%%%%%%%%%%%%%%%%%%%%%%%%%%%%%

\section{Chiral phase transition in the RN background}
\label{sectionVI}

The objective of this work is to examine the chiral phase transition at finite temperature and chemical potential in a nonlinear charged black hole background. First, we can reproduce the behaviour of the order parameter, chiral condensate, in the RN background by taking the large value of $\beta$, which results in the reduction of the BI geometry to RN geometry.

\begin{figure}
\begin{subfigure}{0.48\textwidth}
  \centering
  \includegraphics[width=1\linewidth]{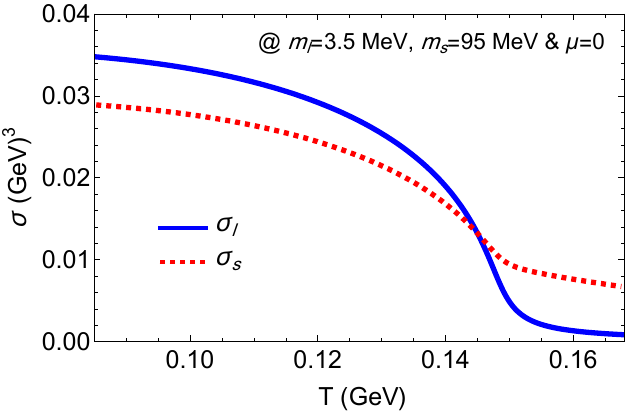} 
   \caption{ }
\label{sigmaphysicaa}
\end{subfigure} 
\begin{subfigure}{0.5\textwidth}
  \centering
  \includegraphics[width=1\linewidth]{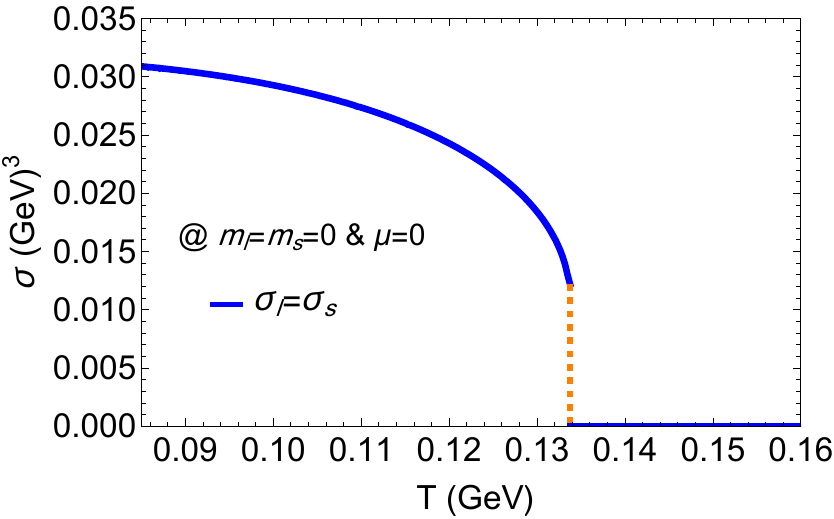}
 \caption{ }
\label{sigmachiralb}
\end{subfigure}
\caption{Panel (a): Temperature dependence of the light quark condensate with 
the large strange quark mass $m_{s}=1$~GeV and charm quark mass $m_{c}=3$~GeV. This case corresponds to the two-quark flavor system for the massless (solid blue line) and physical quark mass (dashed red line) of the light quark. Panel (b): The variation of $\sigma_{l}$ with temperature $T$ with similar color coding as Panel ~(a). }
\label{sigmamu0}
\end{figure}

First, we discuss the chiral phase transition at physical quark masses with three-quark flavors at zero chemical potential. In Fig.~\ref{sigmaphysicaa}, we present the light quark condensate $\sigma_{l}$ and $\sigma_{s}$ as functions of temperature for the three-quark flavor system with physical quark masses. The left Panel illustrates that the chiral condensate $\sigma_{l}$ decreases smoothly with increasing temperature, yet remains nonzero even at high temperatures. This indicates that the holographic QCD model undergoes a chiral crossover, and chiral symmetry is approximately restored at finite temperatures. The chiral crossover is characterized by the pseudocritical temperature, which is determined by the inflection point of the quark condensate, $d^2 \sigma_l/d^2 T \bigl|_{T=T_{\rm pc}}=0$, and is evident as a peak in $d \sigma_l/dT$. The strange quark condensate exhibits similar behavior, with a higher value at high temperatures compared to the light quark condensate. For the three-quark flavor system with physical masses of light and strange quarks, the pseudocritical temperature is determined to be $T_{\rm pc}\bigl|_{\rm hQCD} = 0.1477~{\rm GeV}$. To see the effect of the finite quark mass on the chiral condensate, we calculated the light and strange chiral condensate as a function of temperature at the chiral limit, as shown in Fig.~\ref{sigmachiralb}. Now, the light and strange quark condensate becomes degenerate and behaves differently near the critical temperature of the chiral symmetry restoration. The $\sigma_{l(s)}$ decreases as the temperature increases until reaching a point where a jump appears in $\sigma_{l(s)}$ behavior, signaling the first order phase transition with the critical temperature $T_{\rm c}\bigl|_{\rm hQCD} = 0.1337~{\rm GeV}$. The first order region in our model is in contradict with the LQCD results at the continuum limit \cite{Cuteri:2021ikv}, where the chiral phase transition has been shown to be second order. However, a small region of the first order phase transition at the small quark masses does not rule it out. 

\begin{figure}
  \centering
  \includegraphics[width=0.5\linewidth]{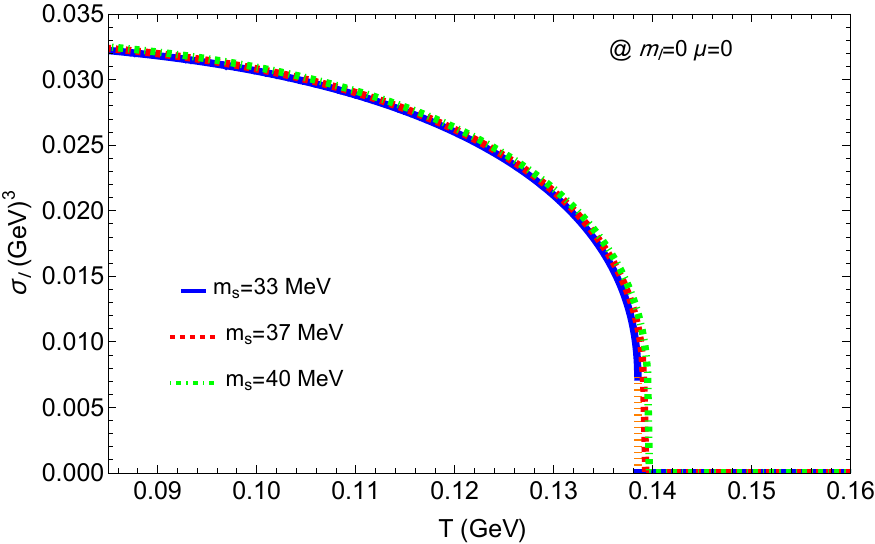} 
\caption{Looking for the critical strange quark mass that changes the order of the chiral phase transition from second order to first order at zero light quark mass and chemical potential.}
\label{sigmamu0ms}
\end{figure}

To see the critical strange mass that changes the order of the phase transition, we fix the light quark mass to be zero and change the strange quark mass as shown in Fig. \ref{sigmamu0ms}. The critical mass that separates the first order domain from the second order region is located at the strange quark mass $m_s=37$ MeV. This strange quark value is below the value that has been used by LQCD collaborations (without taking the continuum limit) to study the order of the chiral phase transition in the $2+1$ flavor system.

\begin{figure}
  \centering
  \includegraphics[width=0.5\linewidth]{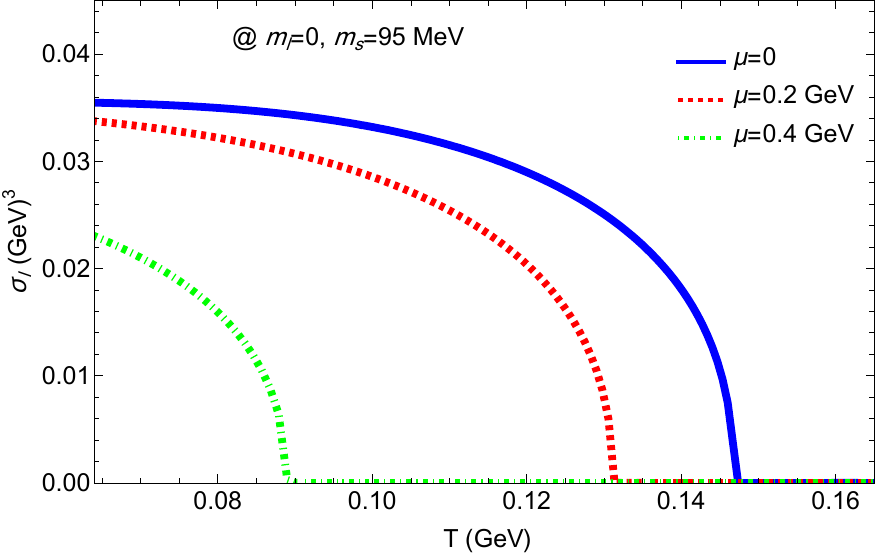} 
\caption{The chiral condensate dependence of the temperature and chemical potential at the quark masses $m_l=0$ and $m_s=95$ MeV.}
\label{sigmaTmu}
\end{figure}

Moving to the case of finite chemical potential, we analyze a $2+1$-flavor system with light quark mass $m_l = 0$ and strange quark mass $m_s = 95\ \text{MeV}$, as illustrated in Fig. \ref{sigmaTmu}. At zero chemical potential ($\mu = 0$), the system exhibits a second-order chiral phase transition with a critical temperature of $T_c = 0.1475\ \text{GeV}$.
Introducing a finite chemical potential of $\mu = 0.2\ \text{GeV}$ preserves the second-order nature of the transition; however, the critical temperature is reduced to $T_c = 0.1314\ \text{GeV}$. Moving to a higher chemical potential of $\mu = 0.4\ \text{GeV}$, the phase transition remains second-order while the critical temperature further decreases to $T_c = 0.089\ \text{GeV}$. Thus, within the examined parameter range, increasing the chemical potential does not alter the order of the chiral phase transition but systematically shifts $T_c$ to lower temperatures, signaling a suppression of chiral symmetry restoration with growing quark density.

\begin{figure}
  \centering
  \includegraphics[width=0.5\linewidth]{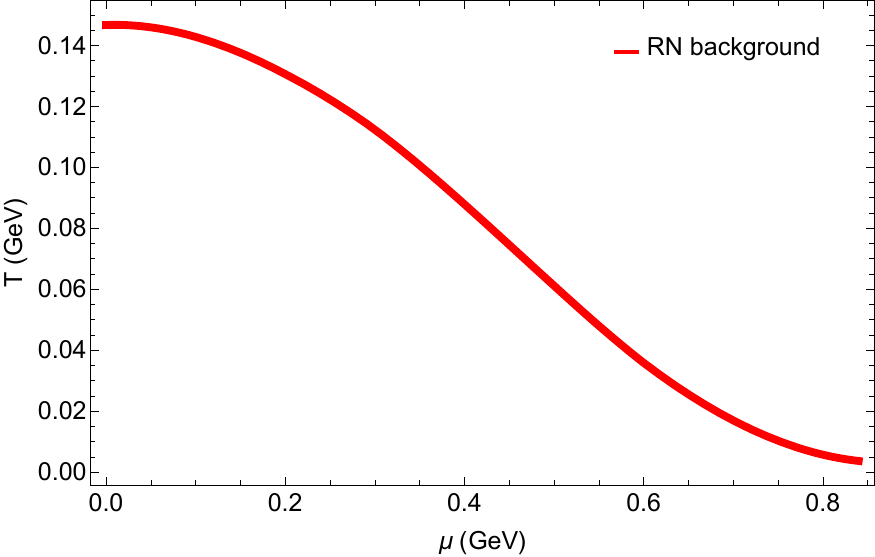} 
\caption{The chiral phase transition in the T-$\mu$ plane. The transition
is second-order with the quark masses $m_l=0$ and $m_s=95$ MeV.}
\label{RNTmu}
\end{figure}

The temperature and chemical potential dependence of the chiral condensate is commonly represented by the chiral phase diagram in the temperature–chemical potential (T-$\mu$) plane, as shown in Fig. \ref{RNTmu}. The solid line in the T-$\mu$ diagram indicates the transition from a chirally broken hadronic phase to a chirally restored phase as temperature or density increases. Notably, the order of the chiral phase transition remains second order throughout the region, and no critical endpoint (CEP) is observed in this model. This result is consistent with previous studies employing the soft-wall model \cite{Colangelo:2011sr,Bartz:2016ufc}. Such behavior is qualitatively consistent with expectations from strongly coupled QCD-like theories and highlights the utility of holography in modeling non-perturbative chiral dynamics under extreme conditions.

\section{Chiral phase transition in Born–Infeld black hole}
\label{sectionVII}

In this section, we present the results for the chiral phase transition within the holographic model constructed in the BI black hole background. The key feature of this setup is the introduction of the nonlinear parameter $\beta$, which controls the strength of the nonlinear electrodynamics in the bulk. As $\beta \rightarrow \infty$, the geometry reduces to the standard RN case. Finite values of $\beta$ modify the near-horizon geometry and the effective gauge coupling, thereby influencing the thermodynamics and chiral symmetry breaking/restoration patterns in the dual boundary theory.

\begin{figure}[htbp]
    \centering
    \includegraphics[width=0.5\textwidth]{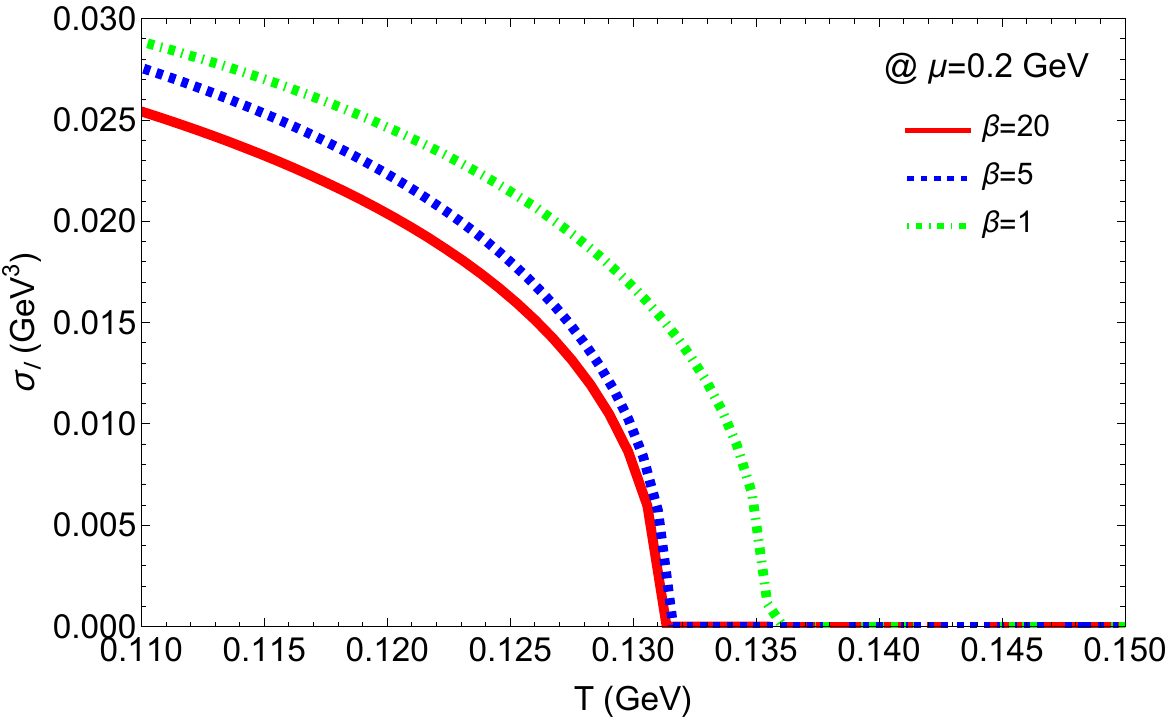}
    \caption{The chiral condensate as a function of temperature $T$ for a fixed chemical potential, $\mu=0.2$ GeV .}
    \label{fig:sigmamu02}
\end{figure}

The behavior of the chiral condensate $\sigma_l$ as a function of temperature $T$ at the fixed chemical potential ($\mu=0.2$ GeV) and various values of $\beta$  is shown in Fig.~\ref{fig:sigmamu02}. The plot shows that for all three values of $\beta$ (1, 5, 20), the order of the phase transition remains intact and the chiral condensate smoothly reaches zero. However, the critical temperature of the chiral symmetry restoration is affected by the value of $\beta$, which increases with decreasing $beta$. Then one can interpret such a result in a way that the nonlinearity of the system delays the melting of the system and chiral symmetry restoration.

\begin{figure}[htbp]
    \centering
    \includegraphics[width=0.5\textwidth]{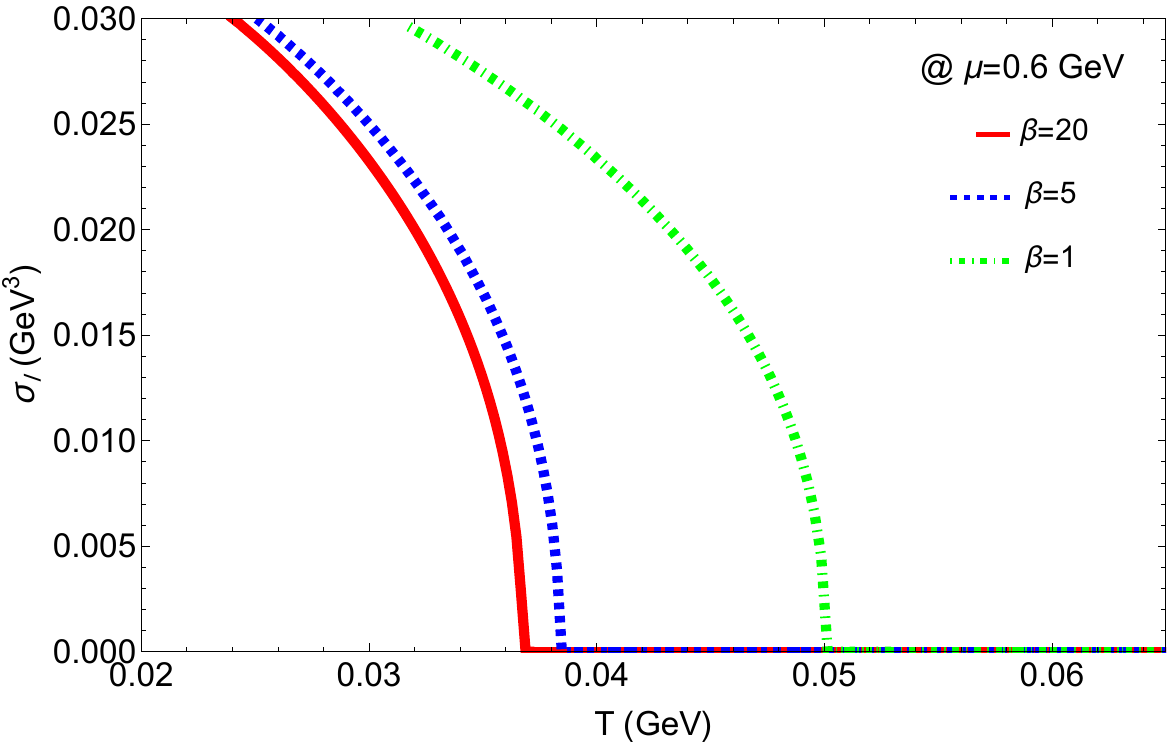}
    \caption{The chiral condensate as a function of temperature $T$ for a fixed chemical potential,$\mu=0.6$ GeV . The gradual decrease indicates a smooth crossover transition.}
    \label{fig:sigmamu06}
\end{figure}

Furthermore, to see the effect of the chemical potential as well, we changed the chemical potential to $\mu-0.6$ GeV in Fig. \ref{fig:sigmamu06}. The qualitative behavior remains consistent for different values of the BI parameter $\beta$, though the value of the critical temperature reduces as expected, reflecting the sensitivity of the chiral phase transition to nonlinear electrodynamic effects.

\begin{figure}[htbp]
    \centering
    \includegraphics[width=0.5\textwidth]{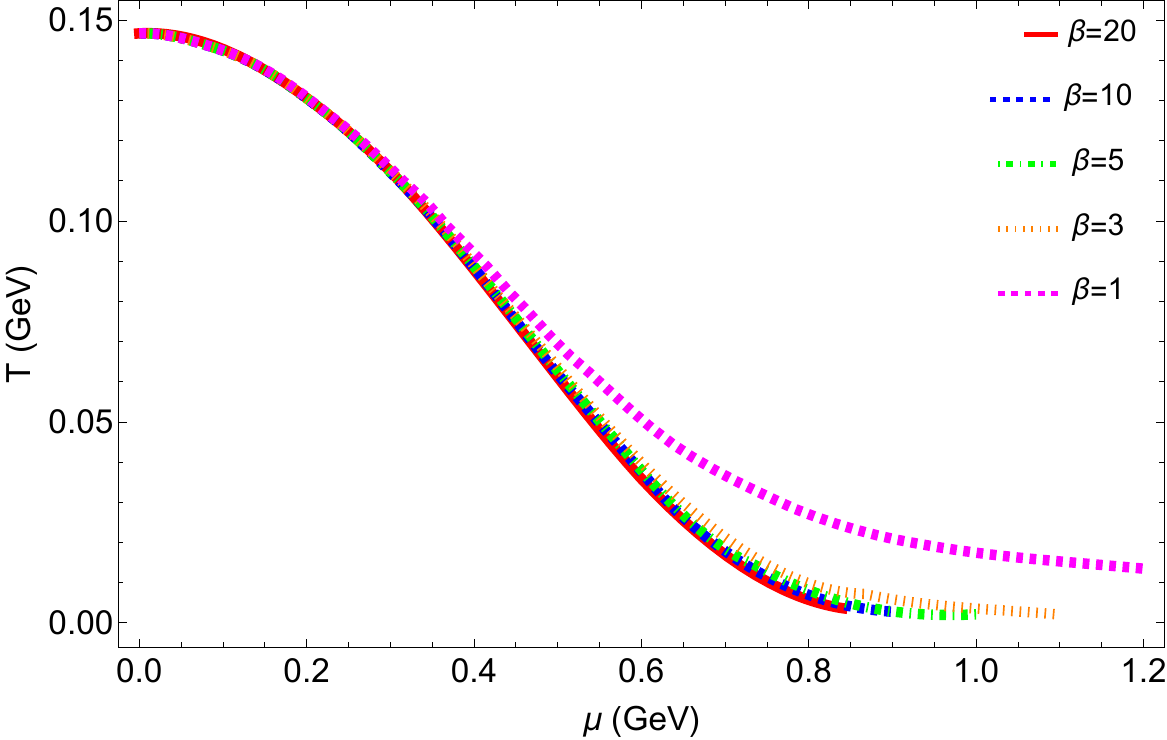}
    \caption{Phase diagram in the $T$--$\mu$ plane for different values of the BI parameter $\beta$. The curve represents the second order phase transition line.}
    \label{fig:Tmu_phase}
\end{figure}

The full phase structure in the temperature-chemical potential plane is illustrated in Fig.~\ref{fig:Tmu_phase}. The diagram delineates the boundary between the chiral symmetry broken phase (low $T$, low $\mu$) and the restored phase (high $T$ and/or high $\mu$).

The primary effect of varying $\beta$ is a systematic shift of the second-order line. As $\beta$ decreases, the phase boundary moves towards higher temperatures at a fixed chemical potential.
In the regime of higher chemical potential, the transition line may exhibit a change in curvature. For sufficiently small $\beta$, the second-order region extends to larger $\mu$, suggesting that strong nonlinearity can suppress the tendency to change the order of the phase transition at low $T$ and high $\mu$.

Over the explored range of $T$ and $\mu$, the chiral transition is a second-order, consistent with LQCD expectations at physical strange quark mass and zero light quark mass and moderate chemical potential.  The BI parameter $\beta$ acts as a tunable scale that modifies the critical dynamics. Stronger nonlinearity (smaller $\beta$) facilitates chiral symmetry restoration at higher temperatures, indicating that nonlinear electromagnetic effects can significantly alter the phase boundary. This demonstrates how incorporating nonlinear electrodynamics in the gravitational dual enriches the phase structure of strongly coupled gauge theories, providing a flexible framework to model chiral dynamics under extreme conditions.

%%%%%%%%%%%%%%%%%%%%%%%%%%%%%%%%%%%%%%%%%%%%%%%%%%%%%%%%%%%%%%%%%%%%%%%%%%%%%%

\section{Chiral Symmetry Restoration via Meson Susceptibilities}
\label{sectionVIII}

To further validate the phase structure observed in the chiral condensate, we analyze the difference between the pseudo-scalar and scalar susceptibilities, $(\chi_{\pi} - \chi_{\sigma})$, as a function of temperature $T$ and chemical potential $\mu$. This quantity serves as a sensitive probe for the restoration of chiral symmetry; the merging of these two channels signals the disappearance of the chiral gap. In the preceding section, we examined the chiral limit for light quarks while maintaining the strange quark mass at its physical value, a configuration that yielded a second-order phase transition. However, the pseudoscalar (pion) susceptibility is proportional to the square of the pion screening mass. Consequently, in the strict chiral limit, the pion susceptibility goes to zero. To circumvent this, we adopt a physical light quark mass of $m_l = 7$ MeV, corresponding to a pion mass of $m_{\pi} \approx 140$ MeV. Under these conditions, the sharp second-order transition is replaced by a smooth crossover, allowing for a precise determination of the pseudocritical temperature through subsequent convergence of the susceptibility difference.

\begin{figure}[htbp]
    \centering
    \includegraphics[width=0.5\textwidth]{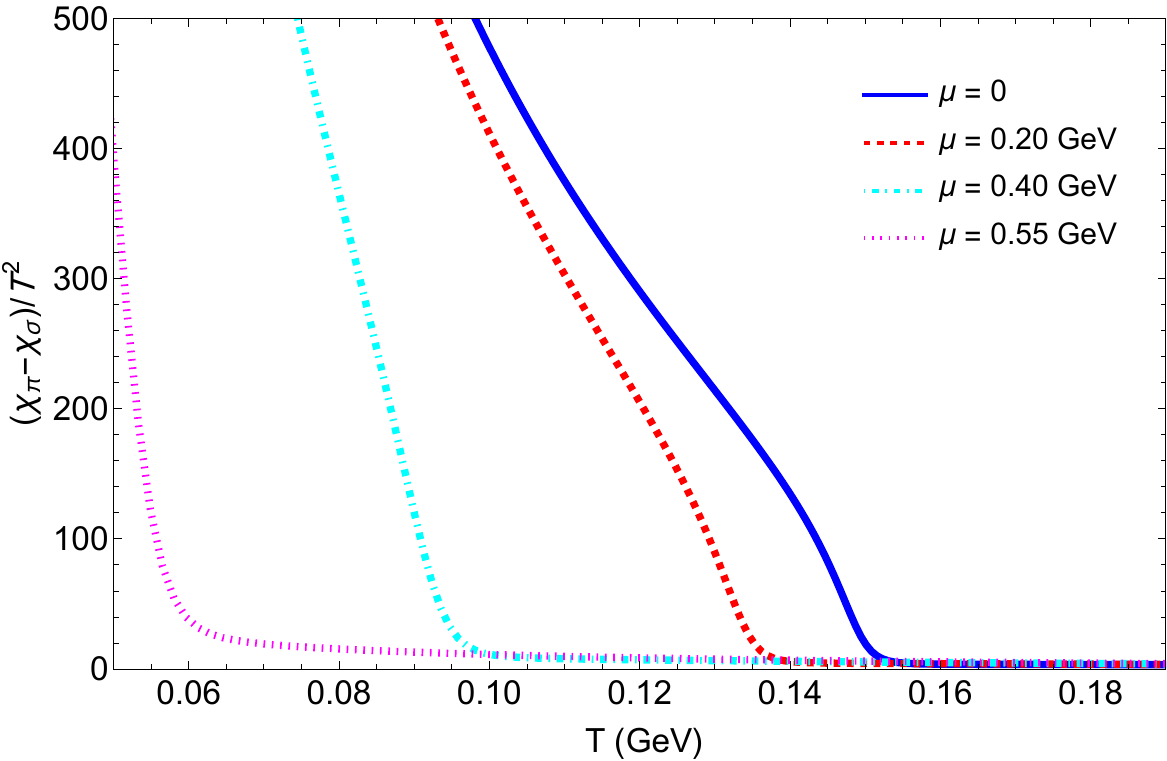}
    \caption{The normalized difference between pseudoscalar and scalar susceptibilities $(\chi_{\pi}-\chi_{\sigma})/T^2$ as a function of temperature $T$ for the RN background. Curves are plotted for chemical potentials $\mu = 0, 0.2, 0.4,$ and $0.55$ GeV.   }
    \label{fig:susRN}
\end{figure}

\begin{figure}[htbp]
    \centering
    \includegraphics[width=0.49\textwidth]{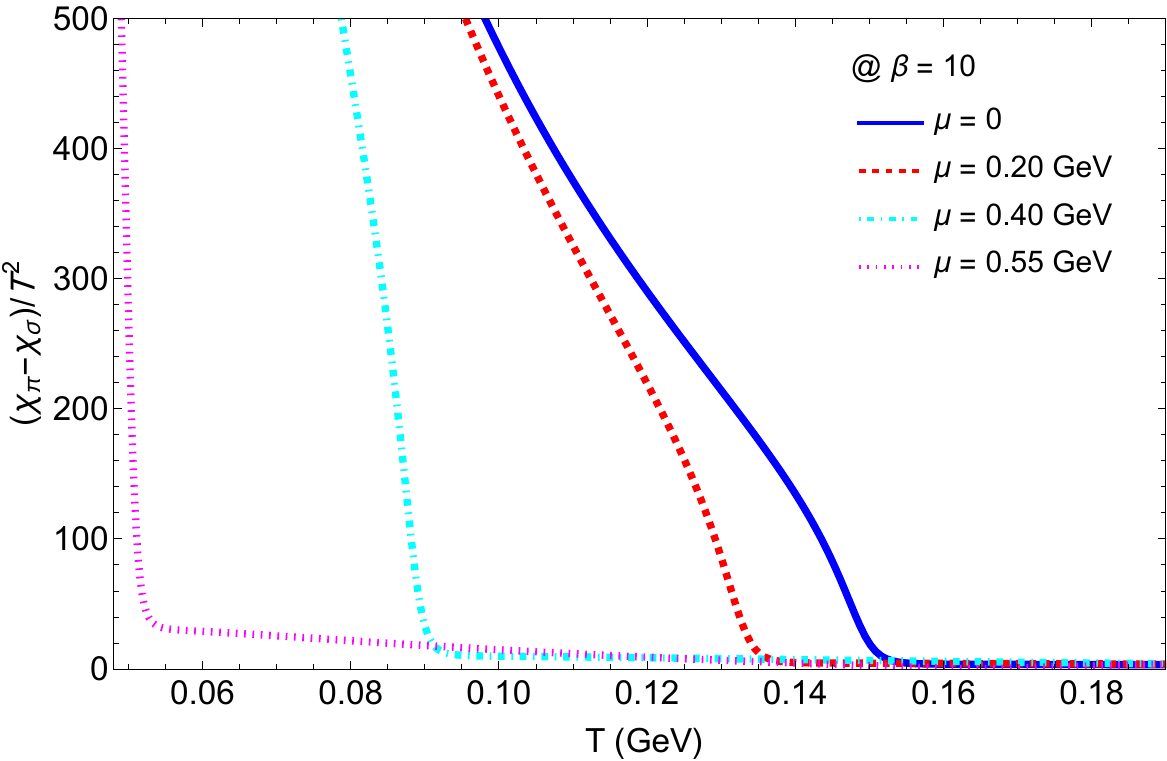}
    \includegraphics[width=0.49\textwidth]{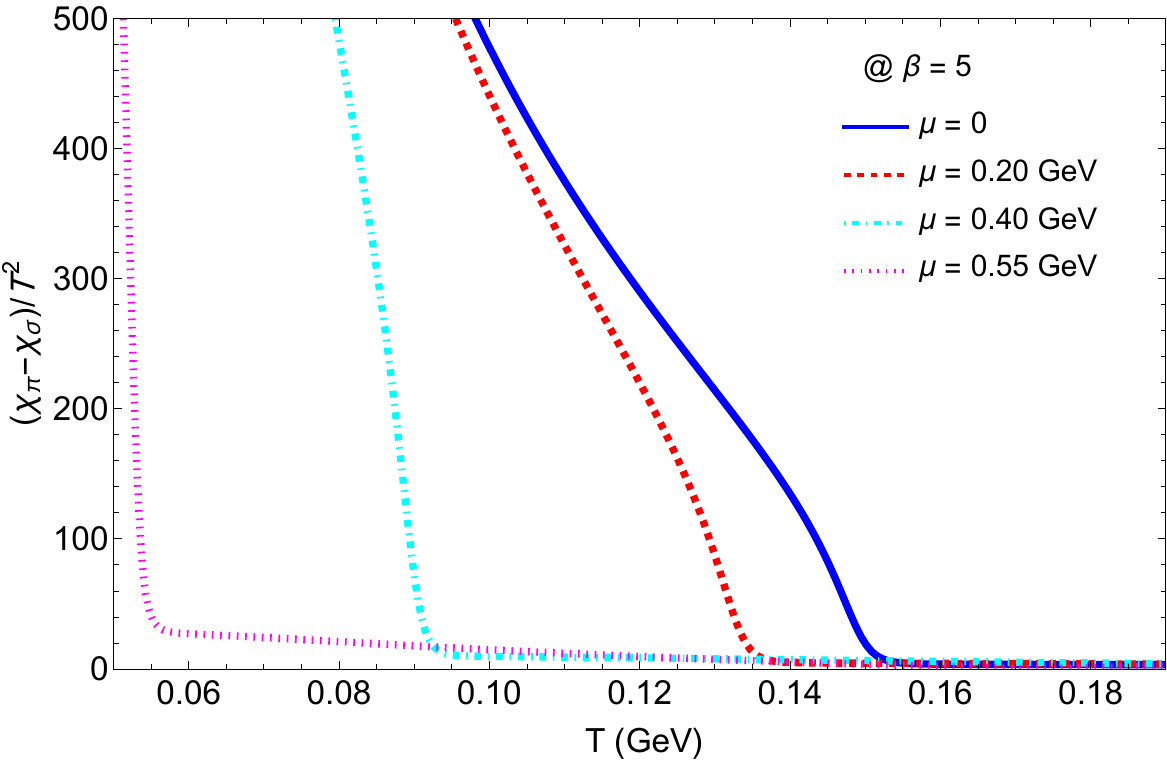}

    \includegraphics[width=0.49\textwidth]{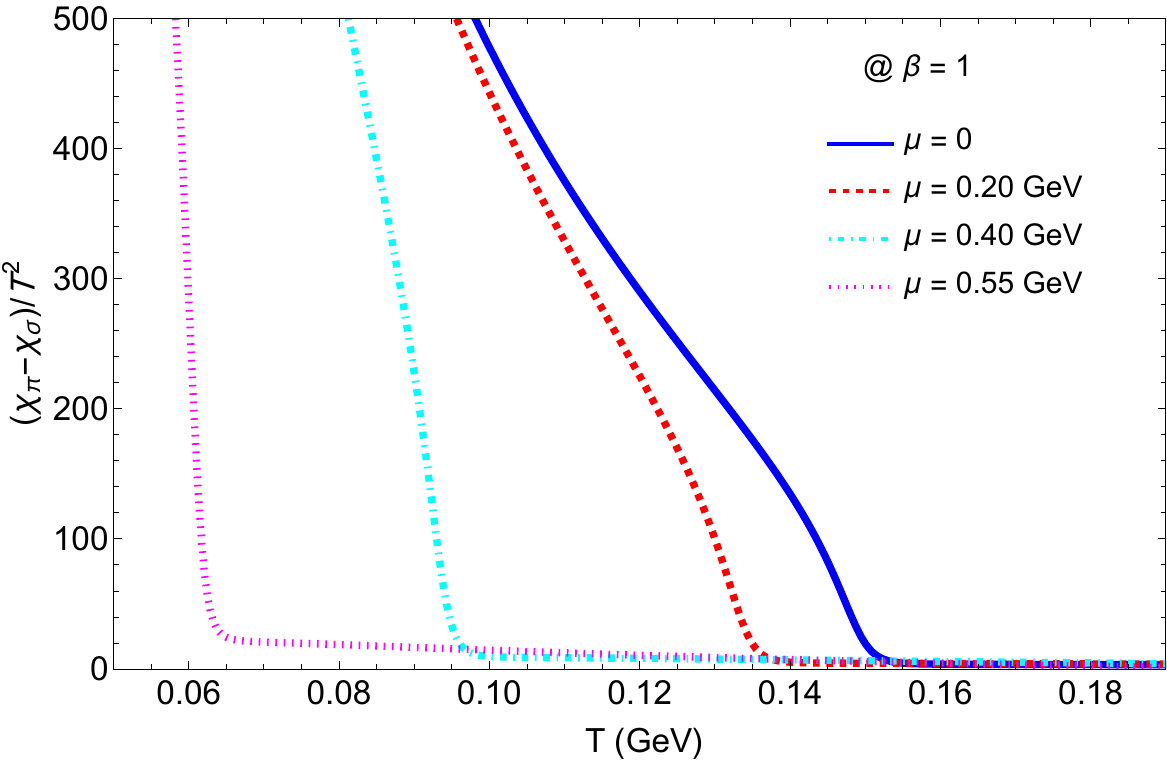}
    \caption{The normalized susceptibility difference $(\chi_{\pi}-\chi_{\sigma})/T^2$ versus temperature $T$ in the BI background with $\beta=10$ (top-left), $\beta=5$ (top-right), and $\beta=1$ (bottom). The rapid decay of the signal indicates chiral restoration, with the transition shifting to lower temperatures as $\mu$ increases.}
    \label{fig:susb}
\end{figure}

Across all considered holographic backgrounds—the RN and the BI black holes—the susceptibilities exhibit a characteristic monotonic decrease with increasing temperature. As shown in Figs. \ref{fig:susRN} and \ref{fig:susb}, the convergence of $\chi_{\pi}$ and $\chi_{\sigma}$ occurs at lower temperatures as the chemical potential $\mu$ increases. This shift confirms that finite density acts as a catalyst for chiral symmetry restoration, reducing the thermal energy required to transition the system into the symmetric phase.

The introduction of the nonlinear parameter $\beta$ provides a significant modification to the phase boundary without altering the second-order nature of the transition. Our analysis reveals several key insights. At $\mu = 0$, the critical temperature $T_c$ remains identical at $0.1474$ GeV for RN, $\beta=10$,$\beta=5$, and $\beta=1$. This indicates that the nonlinear effects of the gauge field are strictly coupled to the charge density of the background. Moreover, at finite chemical potential (e.g., $\mu = 0.55$ GeV), the RN background yields the lowest $T_c$ ($0.0483$ GeV), while $\beta=1$ yields the highest ($0.060$ GeV). A stronger nonlinear electrodynamic effect (smaller $\beta$) effectively stabilizes the chirally broken phase. The BI background modifies the near-horizon geometry in a manner that requires higher thermal energy to achieve the same degree of symmetry restoration compared to the linear Maxwell (RN) case.

The susceptibility results are in perfect alignment with the $T$-$\mu$ phase diagram constructed via the chiral condensate. As the BI nonlinearity is tuned (decreasing $\beta$), the phase boundary is systematically elevated. This consistent behavior across different order parameters reinforces the robustness of our soft-wall model in capturing the interplay between fundamental QCD parameters and the structural nonlinearities of the gravitational dual.

\begin{table} 
\center
\begin{tabular}{ c cc c c}
\hline 
   &  ($T_c$, $\mu_c=0$) GeV   &($T_c$, $\mu_c=0.2$) GeV     &  ($T_c$, $\mu_c=0.4$) GeV    & ($T_c$, $\mu_c=0.55$) GeV \\

\hline
\hline

            RN        &     ($0.1474$ , $0$ )    &  ( $0.1314$ , $0.20$)      &  ($0.0889$ , $0.40$)  &  ($0.0483$ , $0.55$)      \\ 
  
  BI ($\beta=10$)     & ($0.1474$ , $0$ )         &   ($0.1320$ , $0.20$)   &   ($0.0894$ , $0.40$)  &  ($0.0499$ , $0.55$)     \\

  BI ($\beta=5$)          & ($0.1474$ , $0$ )      &  ($0.1321$ , $0.20$)   &   ($0.0897$ , $0.40$)  &  ($0.0510$ , $0.55$) \\

    BI ($\beta=1$)          & ($0.1474$ , $0$ )      &  ($0.1323$ , $0.20$)   &   ($0.0924$ , $0.40$)  &  ($0.0600$ , $0.55$) \\
  
     \hline
     \hline
\end{tabular}
\caption{Critical temperatures $T_c$ (in GeV) for chiral symmetry restoration at various chemical potentials $\mu$ (in GeV). A comparison is shown between the RN background and the BI backgrounds with nonlinear parameters $\beta=10$, $\beta=5$, and $\beta=1$.}
\label{tab:tmu}
\end{table}

%%%%%%%%%%%%%%%%%%%%%%%%%%%%%%%%%%%%%%%%%%%%%%%%%%%%%%%%%%%%%%%%%%%%%%%%%%%%%%

\section{Conclusions}
\label{sectionIX}

We have conducted a comprehensive holographic analysis of chiral symmetry breaking and restoration within a bottom-up AdS/QCD framework, employing nonlinear charged black hole backgrounds. The model successfully implements the holographic dictionary, identifying the chiral condensate $\langle \bar{q} q \rangle$ as a subleading asymptotic coefficient of the scalar field. Numerical solutions for the bulk scalar fields $\chi_f$ confirm stable chiral symmetry breaking in the infrared, with condensates of $\sigma_l \approx 0.272$ GeV$^3$ and $\sigma_s \approx 0.263$ GeV$^3$ at $T=0.100$ GeV for physical quark masses. The character of the chiral phase transition is highly sensitive to quark masses. At zero chemical potential, we observe a smooth crossover for physical masses ($T_{\mathrm{pc}} \approx 0.148$ GeV), and a first-order transition in the chiral limit ($T_{\mathrm{c}} \approx 0.134$ GeV). Keeping the light quark mass at $m_l=0$ and increasing the strange quark mass, we reach a point where the first-order phase transition changes to second order at $m_s=37$ MeV. Beyond this point, the order of the phase transition remains second-order, which is consistent with LQCD \cite{HotQCD:2019xnw}.

To study the effect of the chemical potential on the chiral phase transition, the quark masses are fixed at $m_l=0$, $m_s=95$ MeV. First, we start with the soft-wall model with the RN black hole background (large $\beta$). The critical temperature decreases monotonically with increasing $\mu$, indicating that finite density promotes chiral symmetry restoration.
The $T$-$\mu$ phase diagram constructed within the RN limit shows a second-order transition line without any terminal critical point. The absence of a Critical Endpoint in this setup aligns with prior results in similar soft-wall models.

Our numerical investigation of the chiral phase transition within a BI black hole background reveals that the nonlinearity of the bulk electrodynamics plays a decisive role in stabilizing the chirally broken phase. By systematically varying the BI parameter $\beta$, we have demonstrated that stronger nonlinear effects (smaller $\beta$) push the phase boundary toward higher temperatures across the $T-\mu$ plane. This upward shift in the critical temperature $T_c$ indicates that the nonlinear regularization of the gauge field—which serves as the holographic dual to the quark density—effectively hinders the thermal melting of the chiral condensate.

The analysis of meson susceptibilities $(\chi_{\pi} - \chi_{\sigma})$ provides independent transition markers that perfectly align with the condensate results. Across the RN and BI backgrounds, the susceptibility difference exhibits a rapid thermal decay, signaling the merging of scalar and pseudo-scalar sectors. As density increases, the melting of this difference shifts to lower temperatures, reinforcing the catalytic role of the chemical potential in symmetry restoration. 

The extension to the BI black hole background identifies the nonlinear parameter $\beta$ as a significant modulator of the phase structure. While $\beta$ does not alter the second-order universality class, it systematically shifts the phase boundary: smaller $\beta$ values (stronger nonlinear effects) elevate $T_c$ at fixed chemical potential. This indicates that BI nonlinearity effectively stabilizes the chirally broken phase against thermal restoration. This framework successfully captures how fundamental QCD parameters and the structural nonlinearities of the gravitational dual collectively shape the phase diagram under extreme conditions.

\begin{acknowledgments}

\end{acknowledgments}

\input{main.bbl}% bibliography

%\bibliography{ref}% bibliography

 \addcontentsline{toc}{section}{References}

\end{document}

%% file: main.bbl
%apsrev4-2.bst 2019-01-14 (MD) hand-edited version of apsrev4-1.bst
%Control: key (0)
%Control: author (8) initials jnrlst
%Control: editor formatted (1) identically to author
%Control: production of article title (0) allowed
%Control: page (0) single
%Control: year (1) truncated
%Control: production of eprint (0) enabled
%